\documentclass[amsmath,amssymb,10pt,aps,notitlepage,prl,twocolumn,nofootinbib,longbibliography]{revtex4-1} 
 \usepackage[colorlinks=true]{hyperref}
 \usepackage{lipsum}
 \usepackage{graphicx}
\usepackage{latexsym}
\usepackage{amsmath}
\usepackage{amsthm}
\usepackage{amssymb}
\usepackage{epstopdf} 
\usepackage{enumitem}
\usepackage{setspace}
\usepackage{dcolumn}
\usepackage{bm}
\usepackage{setspace} 
\usepackage{slashed}
\usepackage{color}
\usepackage{youngtab}
\usepackage{tikz}
\usepackage{tikz-3dplot}
\usepackage{braket}
\usepackage{makeidx}
\usepackage{cancel}
\usepackage{subfigure}
\usepackage{pbox}
\usepackage{pifont}
\usepackage{newunicodechar}
\newunicodechar{β}{\Pisymbol{psy}{98}}\newunicodechar{β}{\Pisymbol{psy}{98}}
\usepackage[normalem]{ulem}

\usepackage{lipsum}
\graphicspath{{Figures/}}

\usetikzlibrary{shapes,snakes,arrows,chains,matrix,positioning,scopes,calc}
\usetikzlibrary{decorations.markings}

 \usepackage{moreverb}
\allowdisplaybreaks 
\begin{document}

\title{Construction of Entangled Many-body States via the Higgs  Mechanism }

\author{Pureum Noh}
 \affiliation{Department of Physics, Korea Advanced Institute of Science and Technology (KAIST), Daejeon 34141, Korea}

 \author{Eun-Gook Moon}
\thanks{egmoon@kaist.ac.kr}
\affiliation{Department of Physics, Korea Advanced Institute of Science and Technology (KAIST), Daejeon 34141, Korea}

\date{\today}
\begin{abstract}
We provide a guiding principle to generate entanglement of quantum many-body states by applying key ideas of the Higgs mechanism to systems without gauge structures.
Unitary operators associated with the Higgs mechanism are constructed, named as mean-operators, and employed to prepare entangled many-body states out of a trivial state. We uncover a symmetry-protected-topological state with two Ising symmetries on a square lattice and find entangled states with different symmetries and lattices. 
Plausible applications to quantum simulators such as Rydberg atoms and trapped ions, are also discussed, interpreting the mean-operators as the Ising coupling gates. 
\end{abstract}

\maketitle
{\it Introduction : }  
The Higgs mechanism plays a central role in modern physics of gauge bosons \cite{HM}. 
Mass of gauge bosons is generated by coupling the bosons to low energy excitations such as Goldstone modes from global symmetries. In superconductors, the condensation of the Cooper pairs generate the mass of U(1) electromagnetic gauge fields, explaining the Meissner effect \cite{superconductor}, and the W and Z bosons of the electroweak symmetry become massive by the SU$(2)$ doublet field in the Standard model of particle physics \cite{Higgs}. Extensive research has been done in quantum many-body systems, deepening out understanding in gauge particles \cite{Varma, disordered, Bloch, BJKim}. 

In the Higgs mechanism, the unitary gauge choice plays a special role because decoupling between gauge bosons and matter fields manifests. 
In terms of a superconductor with U(1) electromagnetism, the decoupling is readily revealed in the action,  
\begin{eqnarray}
\mathcal{S} =\int_{x,t} n_s (\partial_{\mu} \theta -  a_{\mu} )^2 + \cdots \,\, \Rightarrow \,\, \mathcal{S}_u= \int_{x,t}  n_s ( a_{\mu} )^2   +\cdots \nonumber
\end{eqnarray}
where $a_{\mu}$ is for the electromagnetic gauge field, $\theta$ is for the phase field of Cooper pairs, $n_s$ is for the superfluid density, and the integration is over space-time. 
In the unitary gauge choice with the subscript $u$, the decoupling between $a_{\mu}$ and $\theta$ is obvious, explaining the Meissner effect of superconductors. 

In this work, we apply key ideas of the Higgs mechanism to systems without gauge structures, demonstrating that the mechanism becomes one useful guiding principle to construct entangled many-body states. 
Our main strategy is to find unitary transformations which decouples between physical degrees of freedom in systems without gauge structures, which correspond to the unitary gauge choice of the Higgs mechanism. 
The construction is done by the two steps. First, consider a system made of two kinds of degrees of freedom, whose coupling mimics the Higgs mechanism. In lattice systems, this step is readily done by considering site and link variables.  
Second, introduce the minimal model and find a unitary operator to decouple the two types of degrees of freedom.
 We name the unitary operator as the mean-operator in connection with the previously proposed mean-operator theory \cite{mot}.
Importance of a global symmetry in our construction is emphasized. The ground state of the constructed Hamiltonian can become trivial unless a proper global symmetry is imposed due to the celebrated Higgs-confinement in gauge theories \cite{Fradkin79}. We note that a similar strategy has been applied to one dimensional systems, for example, by connecting the Kitaev chain with introduced gauge fields, so called a gently gauged model \cite{Sergej}.

One of our main results  is an explicit construction of a symmetry-protected-topological (SPT) state with the global $\mathbb{Z}_2 \times \mathbb{Z}_2$  on a square lattice. Such a SPT state has intriguing entanglement structures \cite{Senthil, Pollman10, LU,spt2,Wen_Chen, classify,spt1,spt3,Levin12,Xu_Ludwig, Xu2014}, and we demonstrate that our construction can be easily extended to systems with different symmetries and lattices by relying on the universal nature of the Higgs mechanism.

{\it The Model : }
To demonstrate our construction, we consider a system with qubits on a square lattice with the periodic boundary condition where site qubits with the Pauli matrices ($\rho^{x,y,z}_i$) live on sites with the index ($i,j$) and link qubits with the Pauli matrices ($\sigma^{x,y,z}_l$) live on links with the index ($l,m$).  The size of the total Hilbert space is $2^{N_{site}+N_{link}}$, and the global symmetry,
\begin{eqnarray}
\mathbb{Z}_2 \times \mathbb{Z}_2= \{ I, \,\, S_{\rho} \equiv \prod_{j} \rho_j^x,\,\, S_{\sigma} \equiv  \prod_{l} \sigma_{l}^x, \,\, S_{\rho} S_{\sigma} \}, \nonumber
\end{eqnarray}
is considered with the identity operator ($I$). The trivial symmetric Hamiltonian and its ground state are, 
\begin{eqnarray}
H_{{\rm triv}} = - \sum_{j} \rho^x_j -  \sum_{l} \sigma_l^x, \,\,|0 \rangle \equiv \big(  \prod_j | \rho_{j}^x = 1 \rangle  \big) \otimes \big(  \prod_{l} | \sigma_{l}^x = 1 \rangle \big).  \nonumber 
\end{eqnarray}
The ground state is a unique product state without entanglement and the symmetry action is independent of the presence of boundaries.  

\begin{figure}[t]
\includegraphics[width=0.48\textwidth]{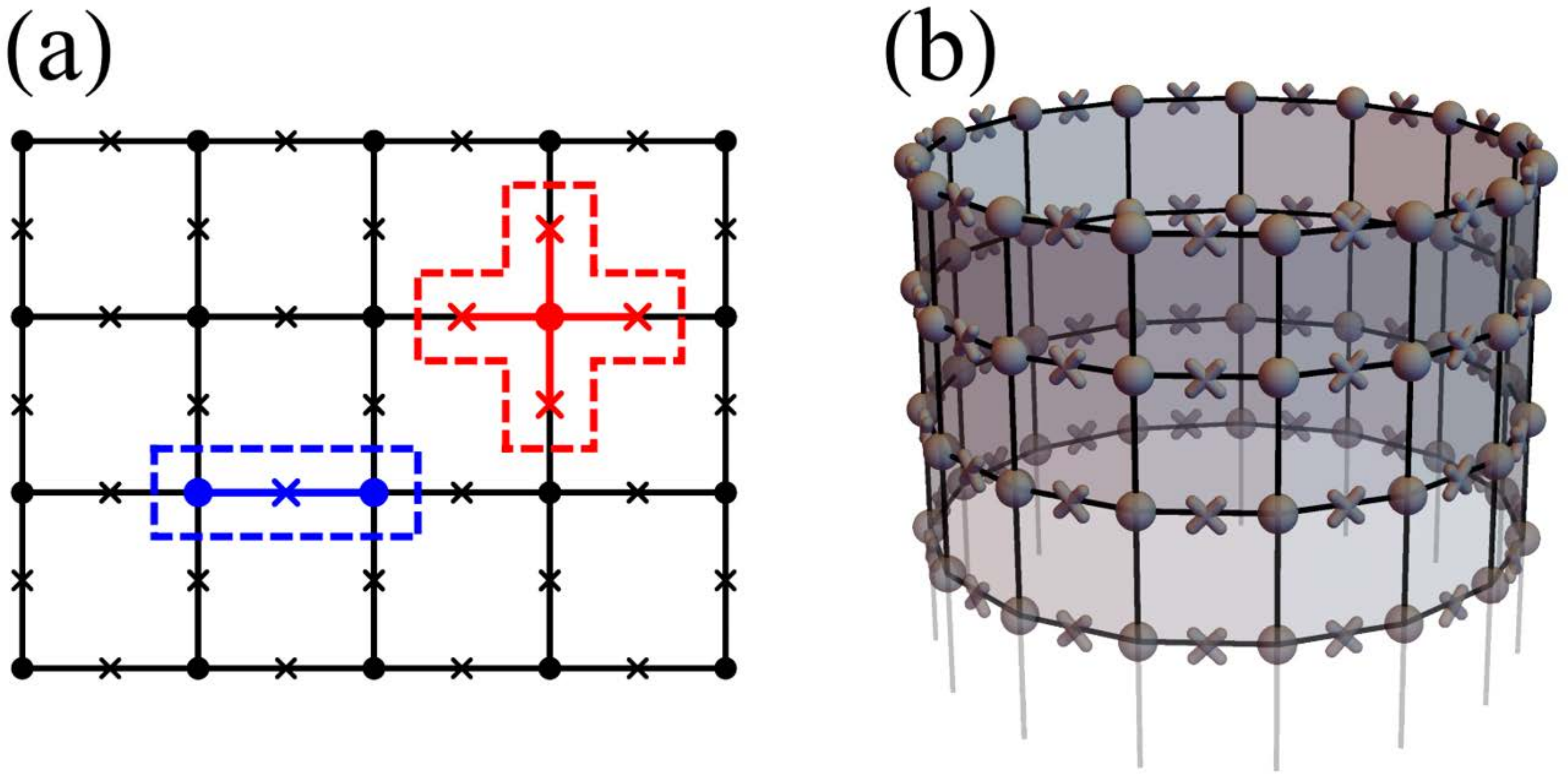} 
 	\caption{ (a) The interaction terms of the minimal model, $H_{{\rm H}} $. The circles (crosses) are for the site (link) qubits, respectively. The blue dotted box is for the Higgs term while the red dotted box is for the Gauss term.  (b) The square lattice on the cylinder geometry.}
 \label{figure}
\end{figure}

We construct an entangled many-body state by finding the ground state of the minimal Hamiltonian,
\begin{eqnarray}
&&H_{{\rm GH}}=  -\sum_j \mathcal{G}_j -\sum_{l}\mathcal{H}_l,\\
{\rm with} \,\,&& \,\,\mathcal{G}_j = \rho_j^x ( \prod_{l \in j} \sigma_l^z), \quad \mathcal{H}_l= \sigma_{l}^x ( \prod_{j \in l} \rho_j^z), \nonumber
\end{eqnarray}
whose graphical representation is given in Fig.\,\ref{figure}(a).
The product in $\mathcal{G}_j$  is over the four links connected to a site $j$, and the product in $\mathcal{H}_l$ is over  the two sites connected to a link $l$. 
We note that generalization to $d$ dimensional hyper-cubic lattices is straightforward. 
The first term, called Gauss term, consists of $2d+1$ qubits, and the second term, called Higgs term, consists of three qubits independent of $d$.

The Hamiltonian enjoys not only the global $\mathbb{Z}_2 \times \mathbb{Z}_2$ symmetry but also the $\mathbb{Z}_2$ local transformation generated by $\mathcal{G}_j$ at a site $j$. 
In spite of the presence of the gauge transformation, we stress that the entire Hilbert space is considered, not the gauge invariant one. 
The effects of the interaction terms such as ($\rho_i^z \rho_j^z$, $\sigma_l^x$) which breaks the local transformation explicitly will be discussed later. 
It is still useful to consider the correspondence. 
The condition of the Gauss term, $\mathcal{G}_j=1$, corresponds to the conventional Gauss law, and the Higgs term describes the Higgs mechanism associated with the $\mathbb{Z}_2$ gauge structure. 
  
The minimal model is exactly solvable since all the terms of $H_{{\rm GH}}$ commute with each other. Its ground state is obtained by finding the mean-operator,  
\begin{eqnarray}
M_{\rm GH}= \big( \prod_{ j,\alpha = \pm \hat{x}, \pm \hat{y}} e^{-i \frac{\pi}{4} \rho^z_j \sigma^z_{j (j+\alpha)}} \big)  \big( \prod_{ l }e^{-i\frac{\pi}{2}\sigma^z_{l}} \big). \nonumber
\end{eqnarray}
The second part of the mean operator, a global $\pi$ rotation of link qubits, is introduced to simplify the notation. Namely, 
the action of the mean operator is precisely mapped to the unitary gauge choice, giving $M_{\rm GH}^{\dagger}( \sigma_{ij}^x \rho_i^{z} \rho^z_{j}) M_{\rm GH} =\sigma_{ij}^x$. However, the presence of the second term is not crucial for construction of entangled states.  
The operator gives the transformation of the Hamiltonian and the ground state,
\begin{eqnarray}
 M^{\dagger}_{\rm GH} H_{{\rm GH}} M_{\rm GH} =H_{{\rm triv}}, \quad |\Psi_{\rm GH} \rangle = M_{\rm GH} |0 \rangle, \nonumber 
\end{eqnarray}
and all excited states and spectrums are easily obtained.

We first show that $| \Psi_{\rm GH} \rangle$ is unique and symmetric under the global $\mathbb{Z}_2 \times \mathbb{Z}_2$ symmetry.  
 The uniqueness is obvious since any qubit flips of $| 0 \rangle$ cost energy. The uniqueness is in sharp contrast to the ground state degeneracy of topologically ordered states. 
The symmetry properties are checked by the relations,
\begin{eqnarray}
S_{\rho,\sigma} |\Psi_{\rm GH} \rangle &=&  M_{\rm GH}^{\dagger} S_{\rho,\sigma} |0 \rangle =|\Psi_{\rm GH} \rangle,\nonumber 
\end{eqnarray}
with the property of the mean-operator, $M_{\rm GH}^2=I$ \cite{mot}. Note that the periodic boundary condition is important in the relation, $M_{\rm GH}^2=I$, which is related to the anomalous symmetry action at boundaries as shown below. 

Next, we investigate entanglement of $| \Psi_G \rangle$ by evaluating the strange correlator, 
\begin{eqnarray}
C^*(i,j)\equiv \frac{\langle  0 | \rho_i^z \rho_j^z| \Psi_{\rm GH} \rangle}{\langle 0  | \Psi_{\rm GH} \rangle} = \frac{\langle 0 | \rho_i^z \rho_j^z  M_{\rm GH} |  0 \rangle}{\langle  0  | M_{\rm GH} | 0  \rangle }, \nonumber 
\end{eqnarray}
which is shown to identify a symmetry-protected-topological state in two spatial dimensions \cite{Xu2014}. 
To evaluate the correlator, we expand the mean-operator,
\begin{eqnarray}
M_{\rm GH} \equiv \sum_{O} \alpha(O) O,  \nonumber
\end{eqnarray}
where  $\alpha(O)$ is a complex number coefficient associated with the operator $O$. Since all the operators contain either $\rho_j^z$ or $\sigma_{l}^z$, we find the relation, $C^*(i,j) = {\alpha(\rho_i^z \rho_j^z)} / {\alpha(I)}$. This procedure is also convenient to see the recently proposed sign-structure of SPT states \cite{sign}, which can be done by comparing signs of two coefficients. 

The key property to evaluate the coefficients is $(\rho_j^z)^2 =(\sigma_l^z)^2 =1$, indicating that only  closed strings of the operators ($\rho^z \sigma^z$) contribute to $\alpha(I)$. Thus, the evaluation is precisely mapped to the  counting problem of  the number of closed strings. Similarly, the other coefficient  $\alpha( \rho_i^z \rho_j^z)$ counts the number of closed strings accompanied with an additional string between the end sites ($i,j$). Thus, the strange correlator is $C^*(i,j)=1$.
Our evaluations can be formulated mathematically, which contain interesting properties of group and set theories (see SM). 

Introducing a physical boundary is another way to find the entanglement structure of $| \Psi_G \rangle$. 
Since the shapes of the boundary depend on how to cut, we choose bulk sites (links) as the sites $j$ (links $l$) to have the four links (two sites) connected to the sites $j$ (links $l$). Our choice guarantees the same number of qubits at the boundary once the translational symmetry of a square lattice is inherited. 

To be specific, let us consider a cylinder geometry as shown in Fig.\,\ref{figure}(b). We focus on how the global symmetry acts on the top edge for simplicity since the inclusion of the bottom edge is straightforward.
We introduce the projection operator ($P$) whose action on quantum states enforces the conditions $\mathcal{H}_l =\mathcal{G}_j=1$ for all $l,j$ in the bulk.
The symmetry actions at the edge with the superscript $(\rm e)$  become
\begin{eqnarray}
&& S_{\rho}^{({\rm e})} =P S_{\rho}P = P\prod_{j \in {\rm edge}} \rho^x_j \sigma^z_{j (j-\hat{y})}P= \prod_{j \in {\rm edge}} \tilde{\rho}^x_j , \nonumber \\
&& S_{\sigma}^{({\rm e})} =P S_{\sigma}P = P\prod_{j \in {\rm edge}} \rho^z_jP= \prod_{j \in {\rm edge}} \tilde{\rho}^z_j, \nonumber
\end{eqnarray}
introducing the edge qubit Pauli operators, $\overline{\rho}^{x,y,z}\equiv PM_{\rm GH}\rho^{x,y,z}M_{\rm GH}^{\dagger}P$.  
Thus, the two Ising symmetries act projectively on the edge qubits with the relations,
\begin{eqnarray}
S_{\rho}^{({\rm e})} &:& \, \overline{\rho}_j^{y,z} \rightarrow -\overline{\rho}_j^{y,z}, \quad \overline{\rho}_j^x \rightarrow +\overline{\rho}_j^x, \nonumber \\
S_{\sigma}^{({\rm e})} &:& \, \overline{\rho}_j^{x,y} \rightarrow -\overline{\rho}_j^{x,y}, \quad \overline{\rho}_j^z \rightarrow +\overline{\rho}_j^z. \nonumber 
\end{eqnarray}
Then, the gapless excitations of the site qubits are protected by the projective representation of the $\mathbb{Z}_2 \times \mathbb{Z}_2$ symmetry at the edge, as shown by the recent development of the Lieb-Schutlz-Mattias theorem with discrete symmetries by Ogata and Tasaki \cite{lsm}.
We note that the translational symmetry of the edges of a square lattice is crucial to protect the gapless excitations, which is a signature of topological crystalline states \cite{Qi}.

{\it Generalization :} Our results on a square lattice are easily generalized by the universal nature of the Higgs mechanism. For a qubit chain, the minimal model transforms to the cluster model  because both of the Gauss and Higgs terms have three qubit interactions. It is well known that the ground state of the cluster model \cite{Wen_QI} is a symmetry-protected-topological state protected by the $\mathbb{Z}_2 \times \mathbb{Z}_2$ symmetry. 
For a hypercubic lattice, the same mean-operators upto the unimportant global rotation factors can be easily constructed. We find that the strange correlator is unity in all dimensions and  the degenerate ground state is protected by the $\mathbb{Z}_2 \times \mathbb{Z}_2$ symmetry (see Supplementary Material (SM)).
Thus, we propose that the ground states are candidate states of symmetry-protected-topological phases with the $\mathbb{Z}_2 \times \mathbb{Z}_2$ symmetry on a hypercubic lattice.  

Generalization to different symmetry groups are also straightforward. To be specific, let us consider the global $\mathbb{U}(1) \times \mathbb{U}(1)$ symmetry and its model Hamiltonian on a square lattice,
\begin{eqnarray}
H_{U(1)} &=&\sum_j G_j^2  -\sum_{i,\alpha=\hat{x},\hat{y}}  \cos (\theta_i -\theta_{i+\alpha} + a_{i(i+\alpha)}), \nonumber
\end{eqnarray}
with the oriented link variables ($E_l, a_l$) and $ G_j = n_j -E_{j(j+\hat{x})} +E_{(j-\hat{x})j} -E_{j(j+\hat{y})} +E_{(j-\hat{y})j}$.  The commutation relations are conventional, $[ n_j,e^{i \theta_j}] = e^{i \theta_j}$ and $[E_l,e^{i a_l}]=e^{i a_l}$, and the symmetry actions are 
\begin{eqnarray}
\mathcal{S}_{\theta} (\alpha) = e^{i \alpha \sum_j n_j}, \quad \mathcal{S}_{E} (\beta) = e^{i \beta \sum_{l} a_l}, \nonumber
\end{eqnarray}
with two real values $\alpha,\beta$.   
The Hamiltonian is exactly solvable and has gapless excitations, which can be shown by introducing the corresponding mean-operator,
\begin{eqnarray}
M_{U(1)} = e^{-i \sum_j \theta_j (-E_{j(j+\hat{x})} +E_{(j-\hat{x})j} -E_{j(j+\hat{y})} +E_{(j-\hat{y})j})}. \nonumber
\end{eqnarray}
We find the transformation, 
\begin{eqnarray}
M_{U(1)}^{\dagger}  H_{U(1)} M_{U(1)} = \sum_j n_j^2  - \sum_{\langle i,j \rangle} \cos(a_{ij}) \equiv H_{0, U(1)}. \nonumber
\end{eqnarray}
The gapless excitation appears because the one of the  $\mathbb{U}(1)$ symmetries suppresses the conventional $\sum_l E_l^2$ term. 
In fact, we find that gapless excitations  even exist with the lower symmetry $\mathbb{U}(1) \times \mathbb{Z}$ as shown in SM. 
Besides the gapless excitations, the presence of anomalous symmetry actions at the boundaries is apparent by the condition $a_{ij} \rightarrow \theta_i-\theta_j$ for the ground state. Namely, both of the conjugate variables ($n_j, \theta_j$) at the boundary transform non-trivially.
Thus, we expect that the ground state of the Hamiltonian describes the gapless systems with symmetry protected edge modes. 
Note that the recent work in one dimensional spin chain finds critical states with topologically protected edge modes between topological phase transitions \cite{Verresen18}. 
The connections with our model and topological phase transitions is an interesting open question.  \\

{\it Discussion and Conclusion : } We stress that the ground state $|\Psi_G \rangle$ respects the Gauss law condition, $\mathcal{G}_j=1$ for all $j$, and the knowledge of well-developed gauge theories is applicable to deepen understanding in  the properties of $|\Psi_G \rangle$. 
We consider the modified minimal model,
\begin{eqnarray}
H = H_{{\rm GH}} - h_{\sigma}\sum_{l} \sigma_l^z -h_{\rho} \sum_{j} \rho_j^z - J \sum_{\langle i,j \rangle} \rho_{i}^z \rho_j^z. \nonumber
\end{eqnarray} 
The terms with two Zeeman couplings ($h_{\rho,\sigma}$) breaks the global $\mathbb{Z}_2 \times \mathbb{Z}_2$ symmetry while the term with $J$ is symmetric.
On the other hand, only the term with $h_{\sigma}$ respects the local $\mathbb{Z}_2$ transformation while the other two terms break. 
It is well known that the term with  $h_{\sigma}$ induces the confinement in the Ising gauge theory, giving the crossover between the Higgs phase and the confined trivial phase \cite{Fradkin79}. In our system, the crossover is prohibited by the global $\mathbb{Z}_2 \times \mathbb{Z}_2$ symmetry, indicating that $|\Psi_G \rangle$  cannot be adiabatically connected to a trivial confined state unless the global symmetry is broken. Similarly, the interaction terms of the other coupling constants can provide a state adiabatically connected to a trivial product state but the global symmetry must be broken explicitly or spontaneously.

Another interesting limit is accessed by turning on the transverse field term of the site qubits, $-\sum_j \rho_j^x$, which is invariant under both the global and local symmetries. Since it does not commute with the Higgs term, quantum fluctuations are introduced and the effects of the Higgs term are suppressed. The suppression of the Higgs term by the transverse field term may provide the route to the $\mathbb{Z}_2$ topological order from the link qubits once the gauge invariant interactions are introduced. This is also consistent with the celebrated Higgs-confinement phenomena \cite{Fradkin79}.   

We propose that the mean-operator associated with the Higgs mechanism is applicable to quantum simulators such as Rydberg atoms and trapped ions by using the Ising coupling gates \cite{ising_gate} along the $z$ direction with the Pauli matriices $Z_{i,j}$, 
\begin{eqnarray}
U(t) = e^{-i t \big( J\sum_{\langle i, j\rangle} Z_{i} Z_{j} \big)}. \nonumber
\end{eqnarray}
The exchange interaction between nearest neighbor qubits are known to be implemented in Rydberg atoms and trapped ions. In experiments with the arrays of Rydberg atoms, the time scale is $t_{{\rm Rydberg}} \approx 0.5\mu$s for large principal quantum numbers ($n\approx 50$) at distances between atoms around 5 $\mu$m \cite{rydbergvalue} while it  is $t_{Trapped}\approx 0.1m$s for trapped ions \cite{trapped_ions}. We remark that our proposal contains two qubits in contrast to the previous proposal with three nearest neighbor qubits \cite{mot} to construct the Levin-Gu states \cite{Levin12}. Manipulations of atom arrays to construct various graphs are already achieved by using Rydberg atoms \cite{rydberg_graph}, and our theory may be applicable to quantum simulators with the hybrid quantum-classical-approximate-optimization \cite{McClean16}. 

In conclusion, we provide a guiding principle to construct entangled states by applying the Higgs mechanism to systems without gauge structures. 
One entangled state with two Ising symmetries on a square lattice is uncovered by showing the presence of protected gapless excitations on edges and calculating the strange correlator, signatures of symmetry-protected-topological states. Generalization to different lattices and symmetries are discussed, and applications to quantum simulators are also proposed. \\

{\it Acknowledgements : } We are grateful for invaluable discussion with Chenjie Wang. 
This work was supported by National Research Foundation of Korea under the grant numbers NRF-2019M3E4A1080411, NRF-2020R1A4A3079707, and NRF-2021R1A2C4001847.

\bibliographystyle{apsrev}
\bibliography{references.bib}
 
\clearpage
\onecolumngrid
\begin{center}
\textbf{\large Supplementary Material for \\``Construction of Entangled Many-body States via the Higgs  Mechanism''} 
\end{center}
\setcounter{equation}{0}
\setcounter{figure}{0}
\setcounter{table}{0}
\setcounter{page}{1}
\section{1.Mathematical Descriptions of mean-operators}
In this section, we discuss properties of a mean-operator which connects a trivial state with a non-trivial symmetry-protected-topological state. 
To be specific, we start with the trivial Hamiltonian and its ground state, 
\begin{eqnarray}
H_{\text{t}} = -\sum_{j}X_j, \quad  |0 \rangle \equiv  \prod_j | X_{j} = +1 \rangle , \nonumber
\end{eqnarray}
with the Pauli matrices ($X_j,Y_j,Z_j$) at a site $j$. 

Considering a mean-operator, $M$, we are interested in the target Hamiltonian, $H_{\text{Target}}=M H_{\text{t}}M^\dagger$, and its ground state, $M|0\rangle$. We impose a  symmetry group, $G=\mathbb{Z}_2^{\otimes n}$, with a positive integer $n$ whose representation uses the Pauli strings of $X_j$. Thus, an eigenstate of the Pauli strings of $X_j$ is symmetric, and the ground state of the trivial Hamiltonian is symmetric. 
We consider a class of mean-operators whose form is expressed as,
\begin{align}
M=e^{-i\sum_a \theta_a A_a},  \label{1.2} 
\end{align}
 where $a$ indexes $Z$-type Pauli strings $A_a$ and real numbers $\theta_a$. We call $A_a$ a \textbf{source} with an index $a$.
 For a generic value of $\theta_a$, the mean-operator breaks $G$, but a certain combination of $\theta_a$ makes the mean-operator symmetric. We define the set of all the sources, \textbf{source set}, $\mathcal{A}=\{A_a| ^\forall a\}$ whose size is $N_{s}$. For a group element $g\in G$, we can consider set $\mathcal{A}_g \equiv \{A_a\in \mathcal{A}|gA_ag^{-1}=-A_a\}$ and $M_g=\prod_{A_a\in \mathcal{A}_g}e^{-i\theta_a A_a}$. Then the following theorem is useful.\\ 
\\
\textbf{Theorem 1.} \textit{ $M|\rm{triv}\rangle$ is symmetric if and only if $M_g^2=I$ for all $g\in G$.} \\
\\
\textit{Proof of Theorem 1:}\\

 \begin{itemize}
\item
($\implies$) For all group elements $g \in G$, $gM_g g^{-1}=M_g^{\dagger}$ and $gM_g^{\dagger}M g^{-1}=M_g^{\dagger}M$  because of the definition of $M_g$.  By acting the group element $g$, we have the relations, 
\begin{align*}
    g \big( M|0\rangle \big) &=\big(g M_g g^{-1}\big)\big(g M_g^{\dagger}Mg^{-1}\big)\big(g|0\rangle\big)= (M_g^{\dagger})^2M|0\rangle=M|0\rangle, \quad M_g^2|0\rangle =|0\rangle.
\end{align*}
Since the trivial state is the linear combination of all configurations of $Z_j$ ($\{z_j\}$), the last equation becomes
\begin{align*}
M_g^2 \big( \frac{1}{\sqrt{2^N}}\sum_{\{z_j\}}| \{z_j\} \rangle \big)=\frac{1}{\sqrt{2^N}}\sum_{\{z_j\}} c({\{z_j\}}) | \{z_j\} \rangle =\frac{1}{\sqrt{2^N}} \sum_{\{z_j\}}| \{z_j\} \rangle,
\end{align*}
with the number of sites, $N$.  The eigenvalue of $M_g^2$ for the eigenstate $|\{ z_j\}\rangle$ is $c({\{z_j\}})$. 
The linear independency of all the eigenstates enforces $ c({\{z_j\}})=1$, which gives the relation, $M_g^2=I$. \\
\item
 ($\impliedby$) For a group element $g \in G$, we have the relations,
\begin{align}
g \big( M|0 \rangle \big) =\big(g M_g g^{-1}\big)\big(g M_g^{\dagger}Mg^{-1}\big)\big( g|0\rangle \big)=\big(g M_g g^{-1}\big) M_g^{\dagger}M|0\rangle. \nonumber
\end{align}
The condition, $M_g^2=I$, gives the relation, $M_g = M_g^{\dagger}$, and thus  $M|\text{triv} \rangle$ is symmetric under $G$. $\Box$\\
\end{itemize} 
If we only consider a class of mean-operators whose sources are all in the same non-trivial irreducible representation of $G$, the following proposition is particularly useful.\\
\\
\textbf{Proposition 1.} \textit{ $M|\rm{triv}\rangle$ is symmetric if and only if $M^2=I$.} \\
 \\
\textit{Proof of Proposition 1:} For all group elements $g \in G$, $\mathcal{A}_g$ are equal to $\mathcal{A}$ or $\varnothing$ because all the sources are in the same non-trivial representation.  It implies $M_{g}$ are equal to $M$ or $I$. Therefore, the condition , $M_g^2=I$ for all group elements $g\in G$, is equal to $M^2=I$ in this case. $\Box$\\
\\
The proposition provides the two-step recipe to construct an entangled symmetric many-body state with $G$, 
\begin{enumerate}
\item  to group the Z-type Pauli strings which are in the same non-trivial representation.
\item  to find values of $\theta_j$ to satisfy $M^2=I$.
\end{enumerate}

Note that the one parameter Hamiltonian, 
\begin{eqnarray}
H (\alpha) = \alpha H_{{\rm t}} + (1-\alpha) M H_{{\rm t}} M^{\dagger}, \nonumber
\end{eqnarray}
satisfies the duality condition, 
\begin{eqnarray}
H(1-\alpha) = M H (\alpha) M^{\dagger}. \nonumber
\end{eqnarray}
At $\alpha=1/2$, we have the commutation relation, $[H(\alpha=1/2),M]=0$, and the self-duality.

We also provide information on how to evaluate $\alpha(O)$ by using group and set theories. Recall that the mean-operator in Eqn. \eqref{1.2} is expressed as 
\begin{align}
    M=\prod_{a} e^{-i\theta_a A_a }=\prod_{a=1}^{N_s}(\cos \theta_a -iA_a \sin \theta_a) \equiv \sum_{O} \alpha(O) O, \label{2.1}
\end{align}  
where a $Z$-type Pauli string $O$ and its coefficient  $\alpha(O)\in \mathrm{C}$ are introduced. The set of complex numbers is denoted as $\mathrm{C}$. We only consider non-trivial $\theta_a$ such that $\cos \theta_a, \sin \theta_a \neq 0$. We consider two sets, $\mathcal{B}\equiv \{O\}$ and $\mathcal{C}\equiv \{\alpha(O)\}$, for all operators $O$ and $\alpha(O)$. The set $\mathcal{B}$ is a group with the matrix multiplication between $Z$-type Pauli strings. The source collection is defined as follows. \\
\\
\textbf{Definition 1 (Source colletion)} \textit{the source collection $\mathbb{S}$ is a collection of all subsets of   $\mathcal{A}$,}
\begin{align*}
 \mathbb{S}&=  \{ \mathcal{S}|\mathcal{S} \subset \mathcal{A} \}.
\end{align*}
\textit{Its element, $\mathcal{S}_{O} \in \mathbb{S}$, is determined by the condition, $\prod_{A_a\in \mathcal{S}_{O}}A_a=O$.}\\
\\
We consider the multiplication $*$ defined as $ \mathcal{S}_{O_1} *  \mathcal{S}_{O_2} \equiv \mathcal{S}_{O_1} \bigcup \mathcal{S}_{O_2}-  \mathcal{S}_{O_1} \bigcap  \mathcal{S}_{O_2}$ for $\mathcal{S}_{O_1}, \mathcal{S}_{O_2} \in  \mathbb{S}$, which makes $(\mathbb{S}, *)$ an abelian group.
Next, let us define a normal subgroup of $\mathbb{S}$, \\
\\
\textbf{Definition 2 (Identity collection)} \textit{the identity collection $\mathbb{I}$ is the collection such that,}
\begin{align*}
 \mathbb{I}&= \{\mathcal{I}\in \mathbb{S}|\prod_{A_a\in \mathcal{I}}A_a=I \}.
\end{align*}
The quotient group $\mathbb{S}/\mathbb{I}$ is naturally introduced by the two definitions. We introduce the homomorphism $f: \mathbb{S} \to \mathcal{B}$ such that $f(\mathcal{S}_{O})=O$, and then its kernel is $\mathbb{I}$. The first isomorphism theorem of the group theory gives that $\mathbb{S}/\mathbb{I}$ is isomorphic to $\mathcal{B}$. This indicates that the mapping, $g: \mathbb{S}/\mathbb{I} \to \mathcal{C}$, such that $g(\mathcal{S}_{O}*\mathbb{I})=\alpha(O)$, exists. The factor map play a key role to find the map $g$. \\
\\
\textbf{Definition 3 (Factor map)} \textit{The factor map $F:\mathbb{S} \to \mathrm{C}$    is defined as}
\begin{align*}
F(\mathcal{S}_{O})=\prod_{A_{a'} \notin \mathcal{S}_{O}}(\cos{\theta_{a'}}) \prod_{A_a \in \mathcal{S}_{O}}(-i\sin{\theta_a}).
\end{align*}
\\
The following proposition and the theorem with the factor map hold. \\
\\
\textbf{Proposition 2.} \textit{For $\theta_a = \theta$, the factor map satisfies the relation, }\\
\begin{align}
F(\mathcal{A}*\mathcal{B})=(\cos{\theta})^{N_s}\frac{F(\mathcal{A})F(\mathcal{B})}{F^2(A\bigcap \mathcal{B})}, \quad \mathcal{A}, \mathcal{B} \in \mathbb{S} .\nonumber
\end{align} \\
\textit{Proof of Proposition 2}: 
The definition of the factor map gives 
\begin{align}
F(\mathcal{A}*\mathcal{B})&=  (\cos{\theta})^{N_s - n(\mathcal{A}*\mathcal{B})}(-i\sin{\theta})^{n(\mathcal{A}*\mathcal{B})}, \quad {\rm for} \quad \theta_a=\theta, \nonumber
\end{align}
with the number of the elements of $\mathcal{A}$, $n(\mathcal{A})$. The relation, $n(\mathcal{A}*\mathcal{B}) = n(\mathcal{A})+ n(\mathcal{B})-2 n(\mathcal{A}\bigcap \mathcal{B}) $, holds.  The factor map becomes
\begin{align}
F(\mathcal{A}*\mathcal{B})&= (\cos{\theta})^{N_s}\frac{(\cos{\theta})^{2n(\mathcal{A}\bigcap \mathcal{B})}}{(\cos{\theta})^{n(\mathcal{A})}(\cos{\theta})^{n(\mathcal{B})}}\times \frac{(-i\sin{\theta})^{n(\mathcal{A})}(-i\sin{\theta})^{n(\mathcal{B})}}{(-i\sin{\theta})^{2n(\mathcal{A}\bigcap \mathcal{B})}} 
=(\cos{\theta})^{N_s}\frac{F(\mathcal{A})F(\mathcal{B})}{F^2(\mathcal{A}\bigcap \mathcal{B})}. \; \Box \nonumber
\end{align}
By definition, the coefficient $\alpha(O)$ is obtained by the relation, $\alpha(O)=\sum_{\mathcal{S}_{O}} F(\mathcal{S}_{O})$. The following theorem gives an alternative way to evaluate $\alpha(O)$. 
\\
\\
\textbf{Theorem 2}. \textit{The coefficient $\alpha(O)$ in $M$ is given by,}
\begin{align}
    \alpha(O)=\sum_{\mathcal{I} \in \mathbb{I}} F(\mathcal{S}_{O}*\mathcal{I}).  \label{thm2}
\end{align}
\\
\textit{Proof of Theorem 2}:  
Considering the two collections,
\begin{align}
 \mathbb{A}=\{\mathcal{S}_{O} |\;\text{for all}\; \mathcal{S}_{O} \in \mathbb{S} \},\quad \mathbb{B}=\{\mathcal{S}_{O}*\mathcal{I}|\;\text{for a fixed}\;\mathcal{S}_{O},\; \text{all}\; \mathcal{I} \in \mathbb{I}\}, \nonumber
\end{align}
we prove the theorem by showing $\mathbb{A}=\mathbb{B}$.

\begin{itemize}
\item
($\mathbb{B}\subset \mathbb{A}$) 
Given $\mathcal{S}_{O}$ and $\mathcal{I}$, the following equations are satisfied, 
\begin{align}
O= \prod_{A_a \in (\mathcal{S}_{O}-\mathcal{I})}\prod_{A_{a'} \in (\mathcal{S}_{O}\bigcap \mathcal{I})}A_aA_{a'},\quad I=\prod_{A_{a''} \in (\mathcal{I}-\mathcal{S}_{O})}\prod_{A_{a'} \in (\mathcal{I}\bigcap \mathcal{S}_{O})}A_{a''}A_{a'}. \nonumber
\end{align}

Consider an element of $\mathbb{B}$ with  $\mathcal{S}_{O}$ and $\mathcal{I}$,
\begin{eqnarray}
\mathcal{B} \equiv  \{ A_a |A_{a} \in (\mathcal{S}_{O}-\mathcal{I})\} \bigcup \{A_{a''}|A_{a''} \in (\mathcal{I}- \mathcal{S}_{O} ) \} \in \mathbb{B}. \nonumber
\end{eqnarray}

Since $A_{a'}^2=I$, we have the relation, 
\begin{eqnarray}
\prod_{A_a \in (\mathcal{S}_{O}-\mathcal{I})}\prod_{A_{a''} \in (\mathcal{I}-\mathcal{S}_{O})}A_aA_{a''} =O. \nonumber
\end{eqnarray}
Thus, $\mathcal{B} \in  \mathbb{A}$. 

\item
($\mathbb{A}\subset \mathbb{B}$) 
Consider two arbitrary elements of $\mathbb{A}$, $\mathcal{S}_{O}$ and $\tilde{\mathcal{S}}_{O}$. The following equations are satisfied, 
\begin{align}
\prod_{A_a \in (\mathcal{S}_{O}-\tilde{\mathcal{S}}_{O})}\prod_{A_{a'} \in (\mathcal{S}_{O}\bigcap \tilde{\mathcal{S}}_{O})}A_aA_{a'}=\prod_{A_{a''} \in (\tilde{\mathcal{S}}_{O}-\mathcal{S}_{O})}\prod_{A_{a'} \in (\mathcal{S}_{O}\bigcap \tilde{\mathcal{S}}_{O})}A_{a''}A_{a'}=O. \nonumber
\end{align}
Since all the operators are the Pauli strings, we have the relation, 
\begin{eqnarray}
O\times O=\prod_{A_a \in (\mathcal{S}_{O}-\tilde{\mathcal{S}}_{O})}\prod_{A_{a''} \in (\tilde{\mathcal{S}}_{O}-\mathcal{S}_{O})}A_aA_{a''}=I,\nonumber
\end{eqnarray}
which gives the element of $\mathbb{I}$
\begin{eqnarray}
\mathcal{I}_0\equiv \{ A_a |A_{a} \in (\mathcal{S}_{O}-\tilde{\mathcal{S}}_{O})\} \bigcup \{A_{a''}|A_{a''} \in (\tilde{\mathcal{S}}_{O}-\mathcal{S}_{O})\} \in \mathbb{I}. \nonumber
\end{eqnarray}
Then, we have the relation, $\tilde{\mathcal{S}}_{O}=\mathcal{S}_{O}*\mathcal{I}_0 \in \mathbb{B}$. 
\end{itemize}
Since $\mathbb{B}\subset \mathbb{A}$ and $\mathbb{A}\subset \mathbb{B}$ are shown, $\mathbb{A}=\mathbb{B}$ and Eqn. \eqref{thm2} are proven. $\Box$ \\
\\
We note that the theorem 2 and proposition 2 are valid only for mean-operator whose all $\cos \theta_a,\sin \theta_a \neq 0$. Therefore, in general case, we should extract the parts where $\cos \theta_a$ or $\sin \theta_a= 0$ before use the theorem and proposition.

\subsection{Strange correlators and sign structures of $H_{\rm GH}$ on hypercubic lattices}
Let us consider the minimal model on a $d$-dimensional hypercubic lattice. 
The corresponding mean-operator with the notation of the main-text is 
\begin{align}
    M_{\rm GH}=\big(\prod_{j}e^{-i\frac{d\pi}{2}\rho^z_j}\big)\big(\prod_{ l }e^{-i\frac{\pi}{2}\sigma^z_{l}}\big)\big(\prod_{j}\prod_{ l \in j}e^{-i\frac{\pi}{4}\rho^z_j\sigma^z_{l}}\big),\quad M_{\rm GH}^{\dagger} H_{\rm GH} M_{\rm GH} = -\sum_j \rho^x_j-\sum_l \sigma^x_l. \nonumber
 \end{align}
where $N$ is the number of sites of a $d$-dimensional hypercubic lattice. We can check that $M_{\rm GH}|{\rm triv} \rangle$ is invariant under $\mathbb{Z}_2 \times \mathbb{Z}_2= \{S_{\rho} \equiv \prod_{j} \rho_j^x,\,\, S_{\sigma} \equiv  \prod_{l} \sigma_{l}^x, \,\, S_{\rho} S_{\sigma} , \,\, I  \}$ by applying the theorem 1. 
We rewrite the operator as 
\begin{eqnarray}
M_{\rm GH} = M  \times \big(\prod_{j}e^{-i\frac{d\pi}{2}\rho^z_j}\big)\big(\prod_{ l }e^{-i\frac{\pi}{2}\sigma^z_{l}}\big), \quad M \equiv \prod_{j}\prod_{ l \in j}e^{-i\frac{\pi}{4}\rho^z_j\sigma^z_{l}} = \frac{1}{2^{dN}} \prod_{j}\prod_{ l \in j}(I-i\rho^z_j\sigma^z_{l}). \nonumber
\end{eqnarray}

We evaluate the strange correlator, $C^*(i,j)={\alpha_{M_{\rm GH}}(\rho_i^z\rho_j^z)}/{\alpha_{M_{\rm GH}}(I)}={\alpha_{M}(\rho_i^z\rho_j^zZ)}/{\alpha_{M}(Z)}$ where $Z\equiv\prod_j (\rho^z_j)^d\prod_l \sigma^z_l$.
The source set and factor map of $M$ are
\begin{eqnarray}
\mathcal{A} = \{\rho^z_j\sigma^z_l| \text{ for all site } j, \text{ link } l\in j\},\; F(\mathcal{S}) = \frac{1}{2^{dN}}(-i)^{n(\mathcal{S})},  \nonumber
\end{eqnarray}
We can set $\mathcal{S}_Z$,
\begin{eqnarray}
\mathcal{S}_{Z} = \{\rho^z_j\sigma^z_{j,j+\hat{x}}| \text{ for all site } j, \text{for all unit vector $\hat{x}$ of the lattice } \},  \nonumber
\end{eqnarray}
and $\mathcal{S}_{\rho^z_i\rho^z_j}$ consists of elements of line connecting $\rho^z_i$ to $\rho^z_j$. Then we note that two relations,
\begin{align}
n(\mathcal{S}_Z)&=n(\mathcal{S}_Z*\mathcal{S}_{\rho^z_i\rho^z_j})=dN,\label{condition1} \\
n(\mathcal{S}_{\rho^z_i\rho^z_j}\bigcap \mathcal{I})&=\text{even} \;\; \text{ for all }\mathcal{I} \in \mathbb{I}. \label{condition2}
\end{align}
From the proposition 2, and Eqn. \eqref{condition1}, \eqref{condition2}, we can show that $F(\mathcal{S}_{\rho^z_i\rho^z_j}*\mathcal{S}_{Z}*\mathcal{I})=F(\mathcal{S}_{Z}*\mathcal{I})$ for all $\mathcal{I}\in \mathbb{I}$. Finally, by using the theorem 2, we can get $C^*(i,j)$,
\begin{align}
C^*(i,j)=\frac{\alpha_{M_{\rm GH}}(\rho_i^z\rho_j^z)}{\alpha_{M_{\rm GH}}(I)}=\frac{\alpha_{M}(\rho_i^z\rho_j^zZ)}{\alpha_{M}(Z)}=\frac{\sum_{\mathcal{I}\in \mathbb{I}}F(\mathcal{S}_{\rho^z_i\rho^z_j}*\mathcal{S}_{Z}*\mathcal{I})}{\sum_{\mathcal{I}\in \mathbb{I}}F(\mathcal{S}_{Z}*\mathcal{I})}=1. \nonumber
\end{align}

To check sign problem, we can set $\mathcal{S}_{\Sigma_j}$ where $\Sigma_j\equiv\prod_{l \in j} \sigma_l$,
\begin{align}
\mathcal{S}_{\Sigma_j}\equiv\{\rho_j^z \sigma^z_l| \text{ for all } l\in j\}
\end{align} 
Again, we can find relations,
\begin{align}
n(\mathcal{S}_{\Sigma_j}*\mathcal{S}_Z*\mathcal{S}_{\rho^z_i\rho^z_j})&=dN\pm2,\label{condition3} \\
n(\mathcal{S}_{\Sigma_j}\bigcap \mathcal{I})&=\text{even} \;\; \text{ for all }\mathcal{I} \in \mathbb{I}. \label{condition4}
\end{align}
From the proposition 2, and Eqn. \eqref{condition3}, \eqref{condition4}, we can show that $F(\mathcal{S}_{\Sigma_j}*\mathcal{S}_{\rho^z_i\rho^z_j}*\mathcal{S}_{Z}*\mathcal{I})=-F(\mathcal{S}_{\rho^z_i\rho^z_j}*\mathcal{S}_{Z}*\mathcal{I})$ for all $\mathcal{I}\in \mathbb{I}$. Finally, by using the theorem 2, a sign problem is checked,
\begin{align}
\frac{\alpha_{M_{\rm GH}}(\rho_i^z\rho_j^z\prod_{l\in j}\sigma^z_l)}{\alpha_{M_{\rm GH}}(\rho_i^z\rho_j^z)}=\frac{\alpha_{M}(Z\rho_i^z\rho_j^z\prod_{l\in j}\sigma^z_l)}{\alpha_{M}(Z\rho_i^z\rho_j^z)}=\frac{\sum_{\mathcal{I}\in \mathbb{I}}F(\mathcal{S}_{\Sigma_j}*\mathcal{S}_{\rho^z_i\rho^z_j}*\mathcal{S}_{Z}*\mathcal{I})}{\sum_{\mathcal{I}\in \mathbb{I}}F(\mathcal{S}_{\rho^z_i\rho^z_j}*\mathcal{S}_{Z}*\mathcal{I})}=-1. \nonumber
\end{align}

\section{2. Entangled States with continuous symmetries}
In this section, we provide detailed information on the models with continuous symmetry. 
First, let us consider a model Hamiltonian on a square lattice,
\begin{eqnarray}
&&H = - J \sum_{j, \alpha=\hat{x},\hat{y}} \cos( \theta_j - \theta_{j+\alpha} +a_{j(j+\alpha)})  +\sum_j ( n_j - (E_{j(j+\hat{x})}-E_{(j-\hat{x})j} +E_{j(j+\hat{y})}-E_{(j-\hat{y})j} ) )^2+ \alpha \sum_{\langle l,m \rangle} (E_l-E_m)^2. \nonumber
\end{eqnarray}
We consider the Hilbert space with the following conditions of the variables, 
\begin{eqnarray}
\theta_j = \theta_j +2\pi, \quad a_{l}=a_{l}+2\pi, \quad E_{l}, n_j \in \mathbb{Z}. \nonumber
\end{eqnarray}
 The commutation relations of the corresponding operators are given
\begin{eqnarray}
&&[ n_j, e^{i \theta_j}] = e^{i \theta_j}, \quad [ E_l,e^{i a_l}]=e^{i a_l}. \nonumber 
\end{eqnarray}
The model enjoys the global $\mathbb{U}(1)\times \mathbb{U}(1)$ symmetry whose actions are
\begin{eqnarray}
\mathcal{S}_{\theta} (\alpha) = e^{i \alpha \sum_j n_j}, \quad \alpha \in \mathbb{R}, \quad \quad \mathcal{S}_{E} (\alpha') = e^{i \alpha' \sum_{l} a_l}, \quad \alpha \in \mathbb{R}. \nonumber
\end{eqnarray} 
Note that  one of the global symmetry forbids the presence of the conventional term, $\sum_l E_l^2$.
Considering the mean operator, 
\begin{eqnarray}
M = e^{-i \sum_j \theta_j (E_{j,\hat{x}}-E_{j-\hat{x},\hat{x}} +E_{j,\hat{y}}-E_{j-\hat{y},\hat{y}} ) }, \nonumber
\end{eqnarray}
the Hamiltonian becomes, 
\begin{eqnarray}
H = M^{\dagger} H_0 M, \nonumber
\end{eqnarray}
with 
\begin{eqnarray}
H_0 = - J \sum_{l} \cos( a_{l}) +\sum_j n_j^2 +  \alpha \sum_{\langle l, m \rangle}  (E_l-E_m)^2. \nonumber
\end{eqnarray}
It is easy to show that the Hamiltonian contains gapless excitations since the number of the conjugate $E$ fields is smaller than the one of the $a_l$ fields.  

If the symmetry is lowered to the global $\mathbb{U}(1)\times \mathbb{Z}$ symmetry, then a model Hamiltonian with the lowered symmetry is  
\begin{eqnarray}
&&H = - J \sum_{j, \alpha=\hat{x},\hat{y}} \cos( \theta_j - \theta_{j+\alpha} +a_{j(j+\alpha)})  +\sum_j ( n_j - (E_{j(j+\hat{x})}-E_{(j-\hat{x})j} +E_{j(j+\hat{y})}-E_{(j-\hat{y})j} ) )^2- \alpha \sum_l \cos(2\pi E_l). \nonumber
\end{eqnarray}
The symmetry  actions are
\begin{eqnarray}
\mathcal{S}_{\theta} (\alpha) = e^{i \alpha \sum_j n_j}, \quad \alpha \in \mathbb{R}, \quad \quad \mathcal{S}_{E}(m) =e^{i m \sum_l a_l}, \quad m \in \mathbb{Z}. \nonumber
\end{eqnarray} 
By using the same mean-operator, we find the relation, 
\begin{eqnarray}
H = M^{\dagger} H_0 M, \nonumber
\end{eqnarray}
with 
\begin{eqnarray}
H_0 =+\sum_j n_j^2  - J \sum_{l} \cos( a_{l}) -  \alpha \sum_l  \cos(2\pi E_l). \nonumber
\end{eqnarray}
Since the site and link degrees of freedom are separable in $H_0$, the ground state is a product state. At each link, the Hamiltonian, 
\begin{eqnarray}
H_{l} = - J  \cos( a_{l}) -  \alpha  \cos(2\pi E_l), \nonumber
\end{eqnarray}
is the Hofstadter Hamiltonian with a fixed flux, which hosts gapless excitations.

\subsection{3. Protected Gapless Edge Modes of $H_{\text{GH}}$ on the Square Lattice}
The most dramatic distinction between the two types of paramagnets is that $H_{\text{GH}}$ has protected gapless edge modes, while $H_{\text{triv}}$ does not. We split $H_{\text{GH}}$ into the bulk part $H^{\text{Bulk}}_{\text{GH}}$ and edge part $H^{\text{Edge}}_{\text{GH}}$ in the cylinder geometry. The bulkt part $H^{\text{Bulk}}_{\text{GH}}$ include only sites which have complete connections with the nearest sites. The edge part $H^{\text{Edge}}_{\text{H}}$ can be any symmetric Hamiltonian
with local interactions which acts on the spins on or near the boundary. We show $PH^{\text{Edge}}_{\text{GH}}P$ have the gapless ground state protected by the symmetry where $P$ is the projection operator of the bulk ground state,
\begin{align*}
P=\left\{\prod_{j}\frac{1}{2} \left(\rho^x_{j}\prod_{l \in j}\sigma_l^z+1\right)\right\} \left\{\prod_{\langle i,j \rangle}\frac{1}{2} \left(\sigma^x_{i,j}\rho_j^z\rho_j^z+1\right)\right\} \left\{\prod_{\text{ghost}}\frac{1}{2}\left(\rho^z_{\text{ghost}}+1\right)\right\}\left\{\prod_{\text{ghost}}\frac{1}{2}\left(\sigma^z_{\text{ghost}}+1\right)\right\},
\end{align*}
by assuming a particular convention that there is a “ghost” spin in the exterior of the cylinder. The summation runs over bulk sites $i,j$.  For degree of freedom in only edge sites in the projected space, we consider  $\widetilde{\rho}_j^{x,y,z}\equiv M\rho_j^{x,y,z}M^{\dagger}$, $\widetilde{\sigma}_l^{x,y,z} \equiv M\sigma_l^{x,y,z}M^{\dagger}$.
From $\widetilde{\rho}_j^{x,y,z}$, $\widetilde{\sigma}_l^{x,y,z}$, the Hamiltonian of bulk part $H^{\text{Bulk}}_{\text{GH}}$, and $P$ can be expressed as,
\begin{align*}
H^{\text{Bulk}}_{\text{GH}}&=-\sum_{j}\widetilde{\rho}_{j}^{x}-\sum_{l}\widetilde{\sigma}_{l}^{x},\;\; P=\left\{\prod_{j}\frac{1}{2}\left(\widetilde{\rho}^x_{j}+1\right)\right\}\left\{\prod_{l}\frac{1}{2}\left(\widetilde{\sigma}^x_{l}+1\right)\right\} \left\{\prod_{\text{ghost}}\frac{1}{2}\left(\widetilde{\rho}^z_{\text{ghost}}+1\right)\right\}\left\{\prod_{\text{ghost}}\frac{1}{2}\left(\widetilde{\sigma}^z_{\text{ghost}}+1\right)\right\},
\end{align*}
where $j,l$ are site and link indices of the bulk parts. The symmetry actions at the edge $S_{\rho}^{({\rm e})},S_{\sigma}^{({\rm e})}$  become, by defining $\overline{\rho}^{x,y,z}_{j}=P\widetilde{\rho}^{x,y,z}_{j}P$ for each edge sites $j$,
\begin{eqnarray}
&& S_{\rho}^{({\rm e})} =P S_{\rho}P = P\prod_{j \in {\rm edge}} \rho^x_j \sigma^z_{j (j-\hat{y})}P= \prod_{j \in {\rm edge}} \overline{\rho}^x_j , \nonumber \\
&& S_{\sigma}^{({\rm e})} =P S_{\sigma}P = P\prod_{j \in {\rm edge}} \rho^z_jP= \prod_{j \in {\rm edge}} \overline{\rho}^z_j. \nonumber
\end{eqnarray}
We can check that behaviors of $\overline{\rho}^{x,y,z}$ under transformation $S_{\rho}^{({\rm e})},S_{\sigma}^{({\rm e})}$ are TABLE \ref{table1}. Note that two Ising symmetries $\mathbb{Z}_2 \times \mathbb{Z}_2$ behave effectively $\pi$-rotations about the three axes for $\overline{\rho}^{x,y,z}$ at the boundary. This symmetries that can be decomposed nontrivial projective on-site representations protect the gapless excitations of the edge site qubits, as shown by the recent result \cite{lsm}.
\begin{table}[tb]
\centering 
\begin{tabular}{c c c c c} 
\hline\hline 
&$I$&$S^{({\rm e})}_\rho$&$S^{({\rm e})}_\sigma$&$S^{({\rm e})}_\rho S^{({\rm e})}_\sigma$  \\[0.5ex] %
\hline 
$\overline{\rho}^{x}$ &$+$&$+$&$-$&$-$\\
$\overline{\rho}^{y}$ &$+$&$-$&$-$&$+$\\
$\overline{\rho}^{z}$ &$+$&$-$&$+$&$-$\\[1ex] 
\hline\hline 
\end{tabular}
\caption{The behaviors of $\overline{\rho}^{x,y,z}$ under transformation $S^{({\rm e})}_{\rho},S^{({\rm e})}_{\sigma}$.}
 \label{table1}
\end{table}

\end{document}